\documentclass[prb,superscriptaddress,amsmath,amssymb,twocolumn,floatfix]{revtex4}
\usepackage{natbib}
\usepackage{graphicx}         % Include figure files
\usepackage{dcolumn}          % Align table columns on decimal point
\usepackage{bm}               % bold math
\usepackage{upgreek}
\usepackage{booktabs} 				% toprule midrule bottomrule
\usepackage{tabularx}
\usepackage{array}
\usepackage[colorlinks=true,urlcolor=blue,breaklinks=true]{hyperref}

\begin{document}

\title
 {Dresselhaus spin-orbit coupling in [111]-oriented semiconductor nanowires}

\author{A.~Bringer}
\affiliation{Peter Gr\"unberg Institut (PGI-1) and JARA-Fundamentals of Future
                  Information Technology, J\"ulich-Aachen Research Alliance,
                  Forschungszentrum J\"ulich, 52425 J\"ulich, Germany}

\author{S.~Heedt}
\altaffiliation{Present address:
                      QuTech, Delft University of Technology, 2628 CJ Delft, The Netherlands}
\affiliation{Peter Gr\"unberg Institut (PGI-9) and JARA-Fundamentals of Future
                  Information Technology, J\"ulich-Aachen Research Alliance, 
                  Forschungszentrum J\"ulich, 52425 J\"ulich, Germany}

\author{Th.~Sch\"apers}
\email{th.schaepers@fz-juelich.de}
\affiliation{Peter Gr\"unberg Institut (PGI-9) and JARA-Fundamentals of Future
                  Information Technology, J\"ulich-Aachen Research Alliance, 
                 Forschungszentrum J\"ulich, 52425 J\"ulich, Germany}

\hyphenation{InAs na-no-wi-re u-sing con-si-der-ing mo-ni-tored
                       na-no-struc-tures na-no-scale spin-tro-nics na-no-elec-tron-ics}
\date{\today}

\begin{abstract}
The contribution of bulk inversion asymmetry to the total spin-orbit coupling is commonly neglected for group III-V nanowires grown in the generic [111] direction. We have solved the complete Hamiltonian of the circular nanowire accounting for bulk inversion asymmetry  via exact numerical diagonalization. Three different symmetry classes of angular momentum states exist, which reflects the threefold rotation symmetry of the crystal lattice about the [111] axis. A particular group of angular momentum states contains degenerate modes which are strongly coupled via the Dresselhaus Hamiltonian, which results in a significant energy splitting with increasing momentum. Hence, under certain conditions Dresselhaus spin-orbit coupling is relevant for [111] InAs and [111] InSb nanowires. We demonstrate momentum-dependent energy splittings and the impact of Dresselhaus spin-orbit coupling on the dispersion relation. In view of possible spintronics applications relying on bulk inversion asymmetry we calculate the spin expectation values and the spin texture as a function  of the Fermi energy. Finally, we investigate the effect of an axial magnetic field on the energy spectrum and on the corresponding spin polarization. 
\end{abstract}
\maketitle

\section{Introduction}

Spin-orbit coupling is an indispensable ingredient when it comes to the realization of Majorana fermions in III-V semiconductor nanowire-based structures.\cite{Leijnse12,Beenakker13} Here, the shift of the energy dispersions due to spin-orbit coupling together with the Zeeman splitting leads to the formation of a helical gap.\cite{Streda03} Partially covering such a nanowire with a superconductor in conjunction with an external magnetic field  results in the formation Majorana fermions close to the boundaries of the superconductor electrodes.\cite{Lutchyn10,Oreg10} As an experimental signature for the existence of these states a differential conductance peak at zero bias was observed.\cite{Mourik12,Das12,Zhang18} With respect to spin-orbit coupling, usually only the Rashba effect is considered.\cite{Bychkov84} Here, the macroscopic electric field at the surface or at an interface leads to a spin splitting of the energy dispersion. By employing a gate electrode this macroscopic electric field can be controlled and by that the strength of the Rashba effect. This property is not only important to tune the helical gap in a semiconductor nanowire\cite{Heedt17} but it is also essential for the functionality of spin electronic devices.\cite{Zutic04}

In addition to the Rashba effect, in zinc blende III-V semiconductors one also finds spin-orbit coupling originating from bulk inversion asymmetry, the so-called Dresselhaus contribution.\cite{Dresselhaus55} Depending on the specific material its strength can even be comparable to the Rashba coupling. In contrast to the Rashba effect, the bulk inversion asymmetry contribution results in a variation of spin-orbit coupling for electrons propagating along different crystal directions.\cite{Winkler03,Winkler04} In fact, by carefully tuning the strength of the Rashba and Dresselhaus contributions, both can fully compensate each other for certain crystallographic directions, resulting in the formation of a persistent spin helix.\cite{Kunihashi09,Walser12,Sasaki14} Generally, for transport in confined systems, such as wire structures, both spin-orbit coupling contributions have to be taken into account to describe the experimental results consistently.\cite{Scheid08,Wenk11} The question arises as to whether the Dresselhaus contribution will support or even generate spin-splitting in systems of reduced dimension. As shown in a recent theoretical study,\cite{Campos18} in nanowires the effect of the Dresselhaus contribution depends on the growth direction. In most cases, zinc blende type III-V nanowires, e.g.\ InAs and InSb nanowires, grow epitaxially along the crystallographic $\left[111\right]$ direction.  In this geometry the electronic bulk states transform according to the double group representation $\Gamma_4$ of $C_{3v}$~[\onlinecite{Parmenter55,Winkler03}] and there is no spin-splitting for momentum $\vec{k}$ along the growth direction ($\vec{k}\parallel \left[111\right]$).\cite{Luo11} However, the contribution is not generally negligible.~\cite{Kokurin15} In fact, by comparing weak antilocalization measurements of InAs nanowires grown in the $\left[111\right]$ direction with theoretical calculations it was shown that spin-orbit coupling due to bulk inversion asymmetry has to be included explicitly.\cite{Kammermeier16,Kammermeier17} In addition, the level spitting due to spin-orbit coupling does not only affect interference effects such as weak antilocalization, it also has a large impact on the g-factor in nanowires.\cite{Winkler17}

We will address the question of what impact Dresselhaus spin-orbit coupling has on the electronic states in InAs and InSb nanowires grown along the $\left[111\right]$ direction. Here, we will focus on Dresselhaus spin-orbit coupling, only. The inclusion of Rashba spin-orbit coupling is straightforward. Assuming a cylindrical quantum well system with electrons confined close to the nanowire surface, the energy spectra and the corresponding level splittings due to the Dresselhaus contribution are calculated by means of a perturbation approach. Moreover, we analyze how the three-fold rotation symmetry of the Dresselhaus Hamiltonian affects the spin texture of the electronic states. Finally, we discuss the effect of an external magnetic field on the energy momentum dispersion and on the spin density.  

\section{Semiconductor Nanowire: Model System}

\begin{figure}[htb]
	\centering
	\includegraphics[width=0.95\linewidth]{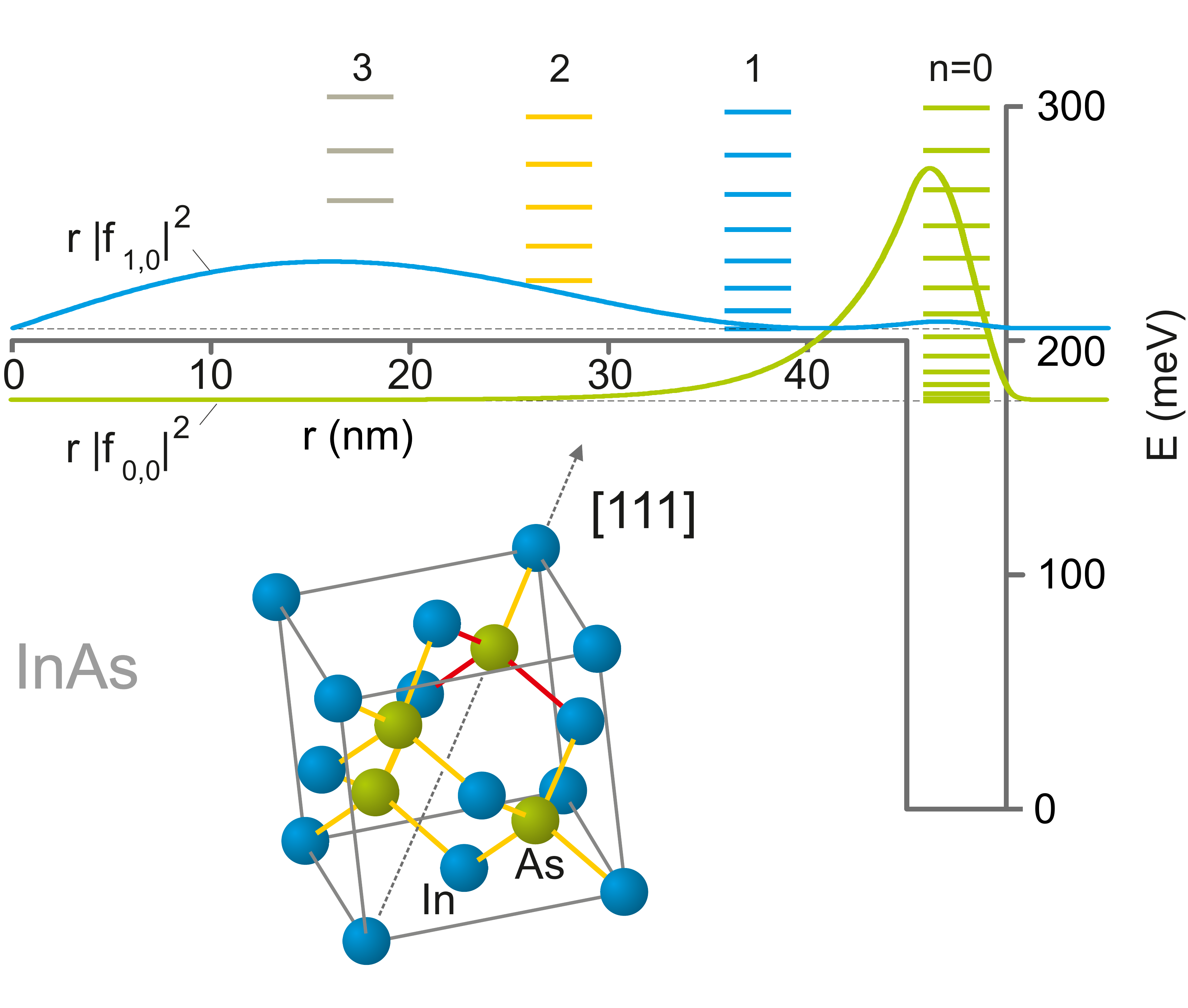}
	\caption[spectrum]
         { Conduction band profile of an InAs nanowire along the radial direction
               with a rectangular $200\,$meV deep quantum well at the outer boundary. The radius of the nanowire is assumed to be $50\,$nm while the quantum well width is $5\,$nm. Also shown are the energy levels $\epsilon_{n,l}$ for $n=0$ to $3$ and the squared amplitude for the states: $n=0, 1$ for $l=0$ without including the Dresselhaus contribution. The inset shows a zinc blende crystal lattice, with the $[111]$ growth direction and the three-fold rotation symmetry (red bonds) indicated.}
	\label{fig:spectrum-InAs-new}
%%%%% Fig. 1
\end{figure}

The dynamics of electrons in cylindrical wires has been discussed already in a previous publication.\cite{Bringer11} There, the confining potential has been calculated self-consistently taking doping effects into account. Here, the Dresselhaus contribution to the energy spectrum and to the wave functions is determined using a perturbation approach. In order to calculate the corresponding matrix elements with acceptable numerical effort and precision, we use a simplified fixed potential profile $V(r)$ with cylindrical symmetry to model the nanowire conduction band. The nanowire radius $r_0$ was set to $50\,$nm, which is a typical value for epitaxially grown InAs and InSb nanowires. As illustrated in Fig.~\ref{fig:spectrum-InAs-new}, we assumed a $5\,$nm wide and $200\,$meV deep rectangular quantum well for the InAs nanowire to mimic the surface accumulation layer due to Fermi level pinning.\cite{Lueth15,Heedt15}
At the surface we assumed a barrier of infinite height. For the InSb nanowire a shallower quantum well with a depth of $100\,$meV and a width of $12.5\,$nm was taken, reflecting the weaker Fermi level pinning at the surface. Electrons confined in the nanowire propagate along the $\left[111\right]$ direction, denoted as the $c$-axis, with a linear momentum $\hbar k$ ($k$ real) and around the axis with an orbital angular momentum $\hbar l$ (integer $l$). The simplified potential profile provides an analytical representation of the wave function  $\psi$
 \begin{equation}
  \psi = e^{ikc} e^{il\varphi} f_{n,l}(r) \; , \label{eq:psi}
 \end{equation} 
 with $c$ the spatial coordinate along the nanowire axis, $\varphi$ the azimuthal angle around the axis, and $f_{n,l}(r)$ the radial distribution function. It solves a radial Schr\"odinger equation
 \begin{equation}
   H_0\,f_{n,l}(r)\,   \,=\,\epsilon_{n,l} f_{n,l}(r) \; , 
 \end{equation}
 with
 \begin{equation}
 H_0\,=\,\frac{\hbar^2}{2m^*}
             \left[-\partial_r^2-\frac{1}{r}\partial_r +\frac{l^2}{r^2}\right]+V\left(r\right). \label{eq:H-null}
\end{equation}
Here, $\partial_r$ is the derivative with respect to the radial coordinate $r$. The quantum number $n=0,1,2, \dots$ counts the nodes of the $f$-functions, which are smoothly connected  Bessel functions for the model potential.
%              Energy
 The energy 
 \begin{equation}
  \varepsilon_{k,n,l} = \frac{\hbar^2}{2m^*}\ k^2 + \epsilon_{n,l} \; ,
  \label{eq:eps0}
 \end{equation}
 consists of the kinetic ($\propto k^2$), the rotational ($\propto l^2$), and the lateral energy $\upsilon_{n,l}$ due to the potential 
 \begin{equation}
  \epsilon_{n,l} = \frac{\hbar^2}{2m^*\bar{r}^2} l^2 + \upsilon_{n,l}.
 \end{equation} 
The proportionality factors are determined by an effective mass $m^*$, i.e.\ $0.026\,m_e$ for InAs and $0.014\,m_e$ for InSb, with $m_e$ the free electron mass. $\bar{r}$ is an average radius determined by the radial distribution function. For electrons in the accumulation layer $\bar{r}$ is close to $r_0$. In Fig.~\ref{fig:spectrum-InAs-new} the energy levels $\epsilon_{n,l}$ for InAs are shown for $n=0$ to $3$ and successively increasing $|l|$. States with $n=0$ and orbital angular momentum up to $l=\pm5$ are energetically located inside the quantum well. Also shown are the squared amplitudes of the radial part of the wave function $r|f_{n,l}|^2$ of the lowest levels for $n=0,1$ at $l=0$. At this point the Dresselhaus contribution has not yet been included. We find, that only the levels with $n=0$ are located close to the surface, while all other states are spread in the whole nanowire volume.
 
In addition to the orbital angular momentum, we also include the electron spin $s=\pm \frac{1}{2}$. Without the Dresselhaus contribution the total angular momentum $j=l+s$ is conserved. The basis functions of the Hilbert space are given by the spinors
 \begin{eqnarray}
  \chi^{\uparrow}_{n,j}&=&
    e^{ikc} e^{il\varphi}f_{n,l}(r) \left(\begin{array}{c} 1\\ 0 \end{array}\right),
   \label{eq:chi-up} \\
  \chi^{\downarrow}_{n,j}&=& 
   e^{ikc} e^{i(l+1)\varphi}f_{n,l+1}(r)\left(\begin{array}{c} 0\\ 1 \end{array}\right).
    \label{eq:chi-down}
 \end{eqnarray}
 The energy eigenvalues scale with the square of the angular momentum quantum
 number $l$. Hence, these two states are not degenerate. 

\section{Dresselhaus Spin-Orbit Interaction}

The wave functions given in Eqs.~\eqref{eq:chi-up} and \eqref{eq:chi-down} are used as a basis to determine the Dresselhaus contribution by a perturbation approach. As a first task, the cartesian Dresselhaus Hamiltonian has to be transformed to a polar coordinate system with an axis along the $[111]$ crystal direction to account for the epitaxial orientation of the nanowire [cf.\ Fig.~\ref{fig:spectrum-InAs-new}(a)]. The Dresselhaus operator for the zinc blende crystal structure is given in $[100]$-orientation by.\cite{Dresselhaus55}
\begin{align}
	H_{\mathrm{D}}	= \gamma_{\mathrm{D}}\left[\right. &\sigma_{x}k_{x}\left(k_{y}^2-k_{z}^2\right)\nonumber\\
	+&\sigma_{y}k_{y}\left(k_{z}^2-k_{x}^2\right)\label{eq:Ham01}\\
	+&\left.\sigma_{z}k_{z}\left(k_{x}^2-k_{y}^2\right)\right]\; , \nonumber
\end{align}
expressed in terms of the Pauli matrices $\sigma_{x,y,z}$,  and the reduced momentum operators  $k_{x} =-i \partial_{x}=-i \partial/\partial x $, and correspondingly for $y$ and $z$. The Dresselhaus coupling parameter $\gamma_{\mathrm{D}}=b_{41}^{6c6c}$ amounts to $27.18\,$eV\AA$^3$\ and $760.1\,$eV\AA$^3$\ for InAs and InSb, respectively.\cite{Knap96,Winkler03} However, recent experimental studies showed that $\gamma_{\mathrm{D}}$ could also be smaller than the theoretical values determined from $k \cdot p$ theory.\cite{Miller03,Faniel11,Walser12,Dettwiler17} In order to account for the growth along the crystallographic $[111]$ direction, we transform a vector $\vec{w}$ represented by
\begin{equation}
\vec{w}=x\hat{e}_x+y\hat{e}_y+z\hat{e}_z,
\end{equation}
corresponding to $[100]$-orientation, to a representation in a rotated basis
\begin{equation}
 \vec{w}=a\hat{e}_a+b\hat{e}_b+c\hat{e}_c \, , 
\end{equation}
 with
\begin{equation}
 \hat{e}_a=\frac{1}{\sqrt{6}}\left(\!\begin{array}{c} 1\\ 1\\ -2 \end{array}\!\right),
   \; \; 
 \hat{e}_b=\frac{1}{\sqrt{2}}\left(\!\begin{array}{c} -1\\ 1\\ 0 \end{array}\!\right), 
   \; \; 
 \hat{e}_c=\frac{1}{\sqrt{3}}\left(\!\begin{array}{c} 1\\ 1\\ 1 \end{array}\!\right),
\end{equation}
 so that $\hat{e}_c$ points along the space diagonal $[111]$ of the original
 unit cell. Hence, the Hamiltonian given by Eq.~\eqref{eq:Ham01} takes the
 following form 
\begin{eqnarray}
H_{\mathrm{D}}	= \gamma_{\mathrm{D}} & \left\{\frac{1}{2\sqrt{3}} \left(k_{a}\sigma_{b}-k_{b}\sigma_{a}\right)\left(k_{a}^2+k_{b}^2-4\,k_{c}^2\right)\right.\nonumber\\
&-\frac{1}{\sqrt{6}}\left[\sigma_{a}k_{c}\cdot 2k_{a}k_{b}+\sigma_{b}k_{c}\left(k_{a}^2-k_{b}^2\right)\right]\nonumber\\
&-\left.\frac{1}{\sqrt{6}}\ \sigma_{c}k_{b}\left(k_{b}^2-3k_{a}^2\right)\right\} \; . 
\label{eq:Ham02}
\end{eqnarray}
 The rotated spin operators $\sigma_{a},\sigma_{b}$, and $\sigma_{c}$
 may be transformed into the Pauli matrices by  a unitary transformation in spin
 space. A useful representation of the spin operators is
\begin{eqnarray}
 \sigma_+ &=& \left(\sigma_{a} + i  \sigma_{b} \nonumber\right)/2 \; ,\\
 \sigma_- &=& \left(\sigma_{a} - i  \sigma_{b}\right)/2 = \sigma_+^{\dagger} 
 \; , 
\end{eqnarray}
 raising and lowering operators with  the properties 
\begin{eqnarray}
 \sigma_+\,\chi^{\downarrow}\, &=&\,\chi^{\uparrow} ,
  \ \sigma_+\,\chi^{\uparrow}\, =\,0 \; ,  \nonumber\\
 \sigma_-\,\chi^{\uparrow}\, &=&\,\chi^{\downarrow} ,
 \ \sigma_-\,\chi^{\downarrow}\, =\,0 \; .  
\end{eqnarray}
With the momenta expressed in cylindrical coordinates $r=\sqrt{a^2+b^2}, \;  \cos\varphi=a/r,\; \sin\varphi=b/r,\; c$\,:
\begin{eqnarray}
 i k_{a}\,=\,\partial_a\,=\,
         \cos\varphi\ \partial_r\ -\ \sin\varphi\ \partial_\varphi/r \; ,\nonumber\\   
 i k_{b}\,=\,\partial_b\,=\,
         \sin\varphi\ \partial_r\ +\ \cos\varphi\ \partial_\varphi/r\; , 
\label{eq:trafo}
\end{eqnarray} 
$H_{\mathrm{D}}$ takes the form (cf.\ Appendix A)
\begin{widetext}
\begin{eqnarray}
 H_{\mathrm{D}} & = &
         \gamma_{\mathrm{D}} \left\{
           \frac{1}{2\sqrt{3}}
            \left[ e^{-i\varphi}\sigma_+
             \left( \partial_{r} - \frac{i }{r}\partial_{\varphi} \right)
                 -  e^{i\varphi}\sigma_-
             \left( \partial_{r} + \frac{i }{r}\partial_{\varphi} \right) \right]
            \left[\nabla^2-5\partial_{c}^2 \right] \right.\nonumber\\
    & & \left. \quad-\frac{i}{\sqrt{6}}
            \left[ e^{2i \varphi}\sigma_+
             \left( \partial_{r}^2-\frac{1}{r}\partial_{r}-\frac{1}{r^2}\partial_{\varphi}^2
               + \partial_{r} \frac{2i}{r}\partial_{\varphi}\right)
               -  e^{-2i\varphi}\sigma_-
               \left( \partial_{r}^2-\frac{1}{r}\partial_{r} -\frac{1}{r^2}\partial_{\varphi}^2
               - \partial_{r} \frac{2i}{r}\partial_{\varphi}\right)
               \right] \frac{1}{i}\partial_{c} \right. \nonumber \\
    & & \left. \quad +\frac{i}{\sqrt{6}}\ \sigma_{c}
            \left[                  \sin\left(3\varphi\right)
             \left( \partial_{r}^3 - \frac{3}{r}\partial_{r}^2
                +\frac{3}{r^2}\partial_{r}
                -\frac{3}{r^2}\partial_{\varphi}^2
                           \left( \partial_{r} - \frac{2}{r} \right) \right)
           \right.\right.\nonumber \\
    & & \left. \left. \qquad\qquad   + \cos\left(3\varphi\right)
           \left(\frac{3}{r}\partial_{r}^2\partial_{\varphi} -\frac{9}{r^2}\partial_{r}\partial_{\varphi}
					+\frac{8}{r^3}\partial_{\varphi} 
						-\frac{1}{r^3}\partial_{\varphi}^3
                          \right) \right] \right\} \; . 
\label{eq:H_D_polar}
\end{eqnarray}
\end{widetext}
The first part of the Hamiltonian in Eqs.~\eqref{eq:Ham02} and \eqref{eq:H_D_polar} conserves the total angular momentum $j$, whereas the second and third parts change the total angular momentum by three (see Appendix A for more details). The three-fold rotation symmetry of the third part is evident in Eq.~\eqref{eq:H_D_polar}, where only the orbital angular momentum is affected, while the second term changes $l$ by $\pm2$ and the spin $s$ by $\pm1$.
 This part of $H_\mathrm{D}$ also has three-fold rotation symmetry. 
 Rotation of $\sigma_+$ by an angle $\alpha$ about the $c$-axis transforms it into 
\begin{equation}
 \exp \left( i \frac{\alpha}{2} \sigma_c \right)
   \sigma_+  \exp\left(-i \frac{\alpha}{2} \sigma_c \right) 
 =  ( \cos \alpha + i  \sin \alpha ) \sigma_+ \; , 
\end{equation} 
 a consequence of the properties of the spin operators:
\begin{equation}
 \sigma_c^2={\bf 1} ,\; \; 
 \sigma_c\sigma_+=\sigma_+ , \; \; \sigma_+\sigma_c=-\sigma_+ \; . 
\end{equation} 
Thus, together with the factor $e^{2 i \alpha}$ for the transformation of the orbital angular momentum, $e^{2 i \varphi}\sigma_+$ transforms into $e^{3 i \alpha} e^{2 i\varphi}\sigma_+$. Similarly, the adjoint operator $e^{-2 i\varphi}\sigma_-$ transforms into  $e^{-3 i \alpha} e^{-2 i \varphi}\sigma_-$. This finally means that a rotation by $\alpha=2\pi/3$ leaves the second line of  Eq.~\eqref{eq:H_D_polar} unchanged. The first line is fully rotational invariant.

\section{Energy Dispersion}

Without the presence of Dresselhaus spin-orbit coupling and at zero magnetic field the states of the nanowire are eight-fold degenerate apart from $l=0$. The energies $\varepsilon_{k,n,l}$ of Eq.~\eqref{eq:eps0} depend only on $\left|k\right|$ and $\left|l\right|$ and are independent of the spin of the electron. Accounting for $H_{\mathrm{D}}$, the eight-fold degeneracy is lifted. Orbital momentum and spin cannot be changed independently, but the matrix elements of $H_{\mathrm{D}}$ are independent of the sign of $j=l+s$. It remains a degeneracy with respect to $\left|k\right|$ and $\left|j\right|$. Each pair of basis states $\chi^{\downarrow,\uparrow}_j$ is linked exclusively to $\chi^{\downarrow,\uparrow}_{j\pm 3}$. Due to the three-fold symmetry  three separated classes of states exist:\\
\begin{equation*}
\begin{array}{ll}
\mbox{class\, "$-1/2$":} & \left\{\chi^{\downarrow,\uparrow}_{n,j}:\, j\,
                        =\ldots,-\frac{7}{2},-\frac{1}{2}, \frac{5}{2}, \ldots   \right\} \; , \\
\mbox{class\, "$1/2$":} & \left\{\chi^{\downarrow,\uparrow}_{n,j}:\, j\,
                        =\ldots,-\frac{5}{2},\frac{1}{2},\frac{7}{2},
                           \ldots  \right\} \; , \\
\mbox{class\, "$3/2$": } & \left\{ \chi^{\downarrow,\uparrow}_{n,j}:\, j\,
              =\ldots,-\frac{9}{2},-\frac{3}{2},\frac{3}{2},
                           \frac{9}{2},\ldots \right\} \; . 
\end{array} 
\end{equation*}
Each state in class "1/2" gets multiplied by $e^{i\pi/3}$ when rotated around the $c$-axis by an angle $2\pi/3$, while for the class "$-1/2$" containing states with reversed sign of $j$ the multiplication factor under rotation is $e^{-i\pi/3}$. In the third class "3/2" the states change sign under rotation. $H_{\mathrm{D}}$ generates stationary states in each class by superposition: 
\begin{equation}
\Psi\,=\,\sum_{n,j,\sigma}\Gamma_{n,j}^{\sigma}\,\chi_{n,j}^{\sigma}, \quad \sigma=\uparrow,\downarrow \, . 
\end{equation}
The coefficients $\Gamma_{n,j}^\sigma$ obey recursion relations (cf.\ Appendix B), which may be solved by "back-folding".

Each $j$ generates a pair of $k$-bands with main component  $l=j\pm 1/2$. The bands start at the energies $\epsilon_{n,l}$ of Eq.~\eqref{eq:eps0} with a small negative shift due to the $k$-independent contributions of the first and the third part of Eq.~\eqref{eq:H_D_polar}. The second part, linear in $k$, is purely imaginary. This leads to matrices in the recursion relations complex conjugate to each other, when the sign of $k$ is changed. The solutions are also complex conjugates -- $\Psi_l(k)^*=\Psi_l(-k)$ -- with equal energy.  $H_{\mathrm{D}}$ is time-reversal invariant. Solutions of class "1/2" and class "$-1/2$" starting from the same $|l|$ at $k=0$ are time-reversed states.  Their energies are equal at $k$ and $-k$. The energy bands of these classes are identical.

\begin{figure*}[ht]
	\centering
	\includegraphics[width=0.7\linewidth]{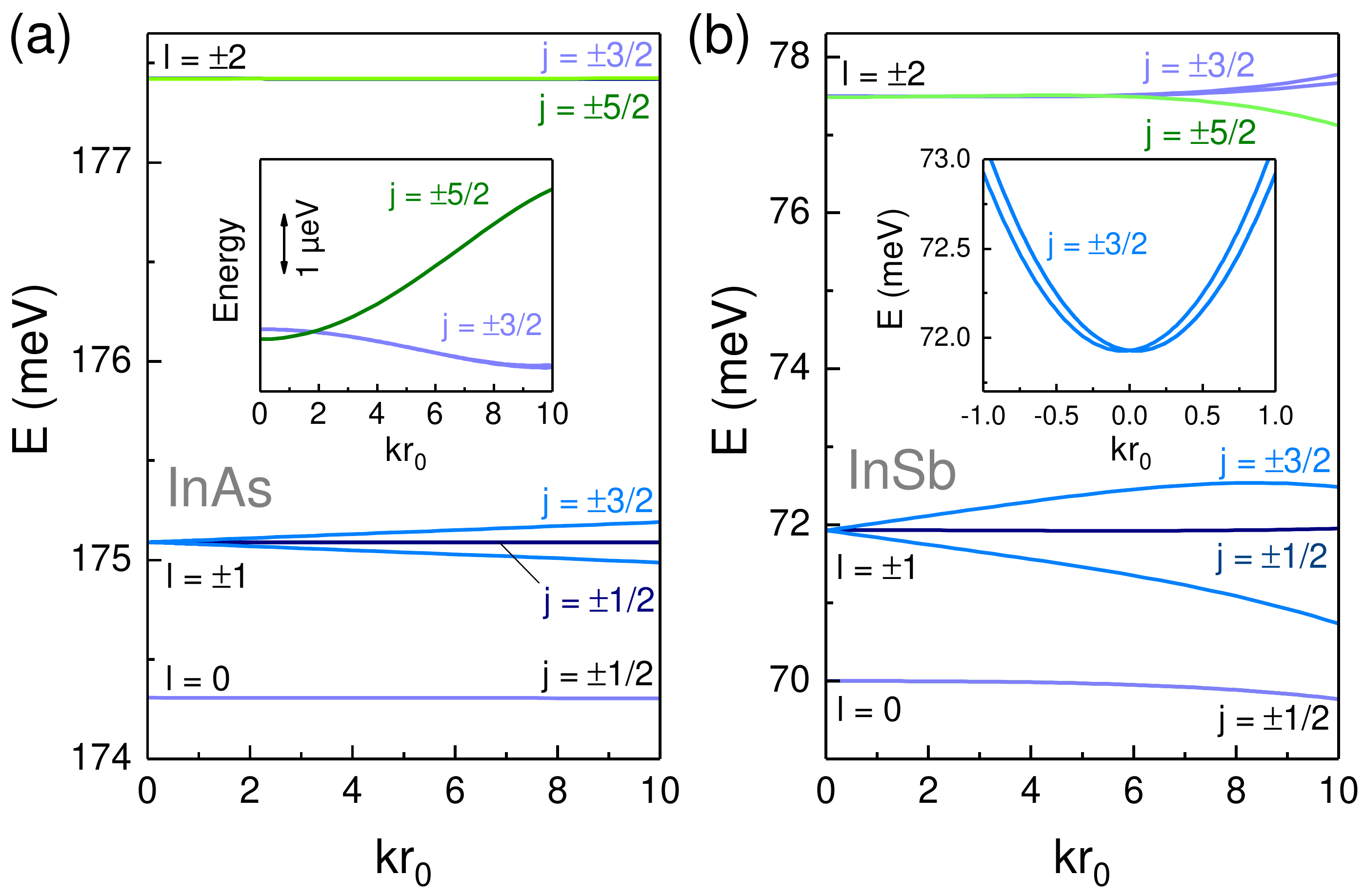}
	\caption[dispersion]{
         (a) Energy vs.\ $k$ dispersion for InAs for an orbital
             angular momentum $l$ from $0$ to $\pm 2$. Note that $l$ refers to
             the dominant contribution to the wave function. The momentum is
             normalized to the nanowire radius $r_0$. The inset shows a detail of the
             upper levels for $l=\pm2$ with $j=\pm \frac{3}{2}$ and
            $\pm \frac{5}{2}$ in a higher energetic resolution.
       (b)~Energy vs.\ $k$ dispersion for InSb nanowires. The inset shows the
            energy vs.\ $k$ dispersion for states with $j=\pm\frac{3}{2}$, $l=\pm1$
            including the kinetic energy for InSb nanowires.}
	\label{fig:InAs-InSb-dispersion}
%%%%%% Fig. 2
\end{figure*}

The class "3/2" contains states with reversed sign of $j$. There is an interaction between them. At $k=0$ the energies of $\Psi_l^\uparrow$ and $\Psi_{-l}^\downarrow$ are equal because of time-reversal invariance, but for finite $k$ there are two energetically different solutions. The two bands cross at $k=0$. Analysis of the recursion relations yield  for the coefficients $\Gamma_{n,j}^\sigma$, $\tilde{\Gamma}_{n,j}^\sigma$ of time-reversed states $\Psi$ and $\tilde{\Psi}$:
\begin{equation}
\tilde{\Gamma}_{n,j}^\downarrow\,=\,\left(\Gamma_{n,j}^\uparrow\right)^* \, , \quad \tilde{\Gamma}_{n,j}^\uparrow\,=\,-\left(\Gamma_{n,j}^\downarrow\right)^* \; .
\end{equation}
The corresponding energy-momentum dispersions are given in
Figs.~\ref{fig:InAs-InSb-dispersion}(a) and (b) for InAs and InSb, respectively. The momentum was normalized by the nanowire radius $r_0$. Furthermore, the contribution from the kinetic energy was omitted. 
Obviously, the modulation of the dispersion in the range up to $kr_0=10$, as shown in Fig.~\ref{fig:InAs-InSb-dispersion}, is weak. For the degenerate $j=\pm \frac{1}{2}$ states with $l=0$ the energy decreases by less than $4\,\mu$eV and by about $230\,\mu$eV for InAs and InSb, respectively. The larger value for InSb is due to the much larger Dresselhaus coupling parameter. In Fig.~\ref{fig:InAs-InSb-dispersion}(a), inset, a detail of the upper levels corresponding to $l=\pm2$ is shown. Owing to the effect of the Dresselhaus contribution, the degeneracy for $j=\pm 3/2$ and $\pm 5/2$ is lifted at $k=0$. At about $kr_0=2$ the levels cross without any hybridization, since the states are from different classes and thus do not couple. A rather large dispersion of the energy is generated by $H_\mathrm{D}$ for the states of class "3/2" with main component $l=\pm 1$. In this case the degenerate components $l=\pm 1$ are coupled directly by the second term of Eq.~\eqref{eq:H_D_polar} which depends linearly on $k$. This leads to two spin-compensated states for $k\neq 0$. The energy splitting increases linearly up to 0.10\,meV for InAs and $0.92$\,meV for InSb at $kr_0=5$ before the quadratic term of Eq.~\eqref{eq:H_D_polar} changes the linear shape of the dispersion. The energy splitting depends linearly on $\gamma_D$. Therefore, for smaller values of $\gamma_D$, as sometimes experimentally observed,\cite{Miller03,Faniel11,Walser12,Dettwiler17} the energy splitting is reduced accordingly. The energy splittings are in the same order of magnitude as the Rashba energy separations observed for nanowires with cylindrical geometry, i.e. for a Rashba coefficient $\alpha_\mathrm{R}$ between $10^{-12}$ and $10^{-11}$\,eVm for InAs we estimated an energy splitting in the range between $0.07$\,meV and $0.7$\,meV for $l=\pm 1$ at $kr_0=5$.\cite{Bringer11} However, one should keep in mind that in the presence of Rashba spin splitting all states are doubly degenerate in contrast to the $j=\pm 3/2$ states discussed here, which are nondegenerate. As shown for InSb in Fig.~\ref{fig:InAs-InSb-dispersion}(b), inset, the minima of the energy bands (including the kinetic energy) are shifted away from $kr_0=0$. For InSb this shift amounts to $\pm 0.043$, while for InAs we find a value of $\pm 0.008$.
\begin{figure}[htb]
	\centering
	\includegraphics[width=0.99\linewidth]{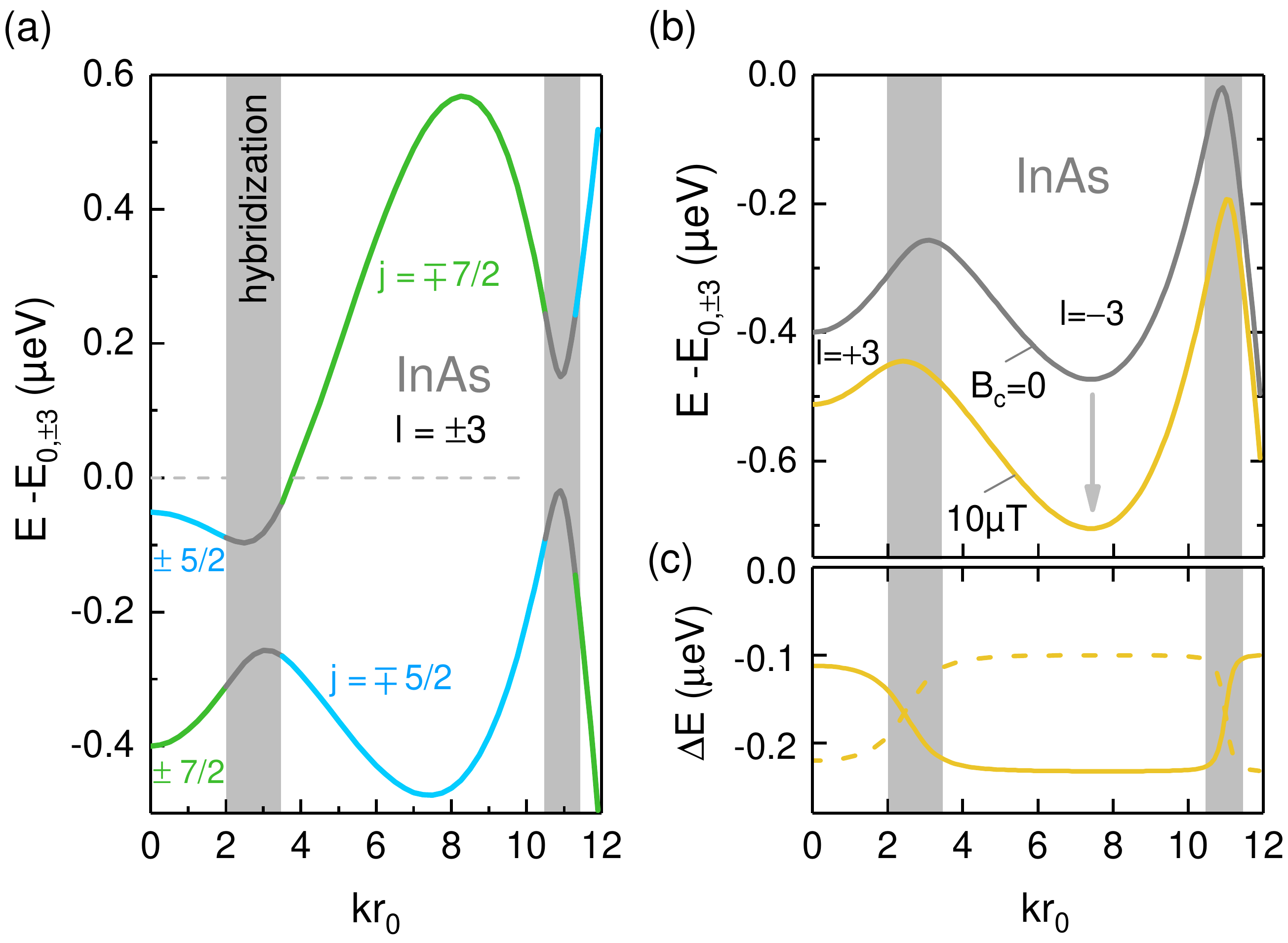}
	\caption[dispersion]
	{(a) Energy vs.\ $k$ dispersion for InAs for the states with $j=\pm      5/2$ and $\pm 7/2$ origination for states with orbital angular momentum $l=\pm 3$. The line color indicate the assignment to $|j|$. Around $kr_0=3$ and $11$ the states hybridize, which results in an avoided crossing (gray shaded areas). (b) Energy vs. $k$ dispersion of the bands starting with $j=+7/2$, i.e.\ $l=+3$ and $s=+1/2$, at $B_c=0$ (gray curve) and at $10\,\mu$T (yellow curve). The ranges of state hybridization are indicated by the gray shaded areas. (c) Energy difference $\Delta E= E(B_c=10\,\mu\mathrm{T})-E(B_c=0)$. The dashed line corresponds to the energy difference for the corresponding state for $s=+1/2$ starting with $l=-3$ at $k=0$.}
	\label{fig:InAs-dispersion-l-3-helical}
\end{figure}
%%%%% Fig. 3
     
Let us return to states of the first two classes. Here, an interesting effect occurs for the two states of class "1/2" with total angular momenta $j=\frac{7}{2}$ and $j=-\frac{5}{2}$, which originate from the initially degenerate states with $l=\pm 3$. As can be seen in Fig.~\ref{fig:InAs-dispersion-l-3-helical}(a), this degeneracy at $kr_0=0$ is removed by the Dresselhaus interaction. Even more, around $kr_0=3$ the two components hybridize strongly. An avoided crossing occurs in conjunction with a rapid change of the angular momentum from $j=\frac{7}{2}$ to $-\frac{5}{2}$ and vice versa. Another hybridization is found at about $kr_0=11$, where the angular momentum is switched back to the initial value at zero momentum. The very same behavior is found for the corresponding states of the "$-1/2$" class with $j=-\frac{7}{2}$ and $\frac{5}{2}$, so that a twofold degeneracy is found, as indicated by the labels in Fig.~\ref{fig:InAs-dispersion-l-3-helical}(a). Since the orbital momentum $l$ is reversed whereas the spin is preserved while passing the hybridization range, the paramagnetic energy shifts are also affected upon application of a magnetic field, as discussed in detail below.     

\section{Spin Texture}

\begin{figure}[ht]
	\centering
\includegraphics[width=0.95\linewidth]{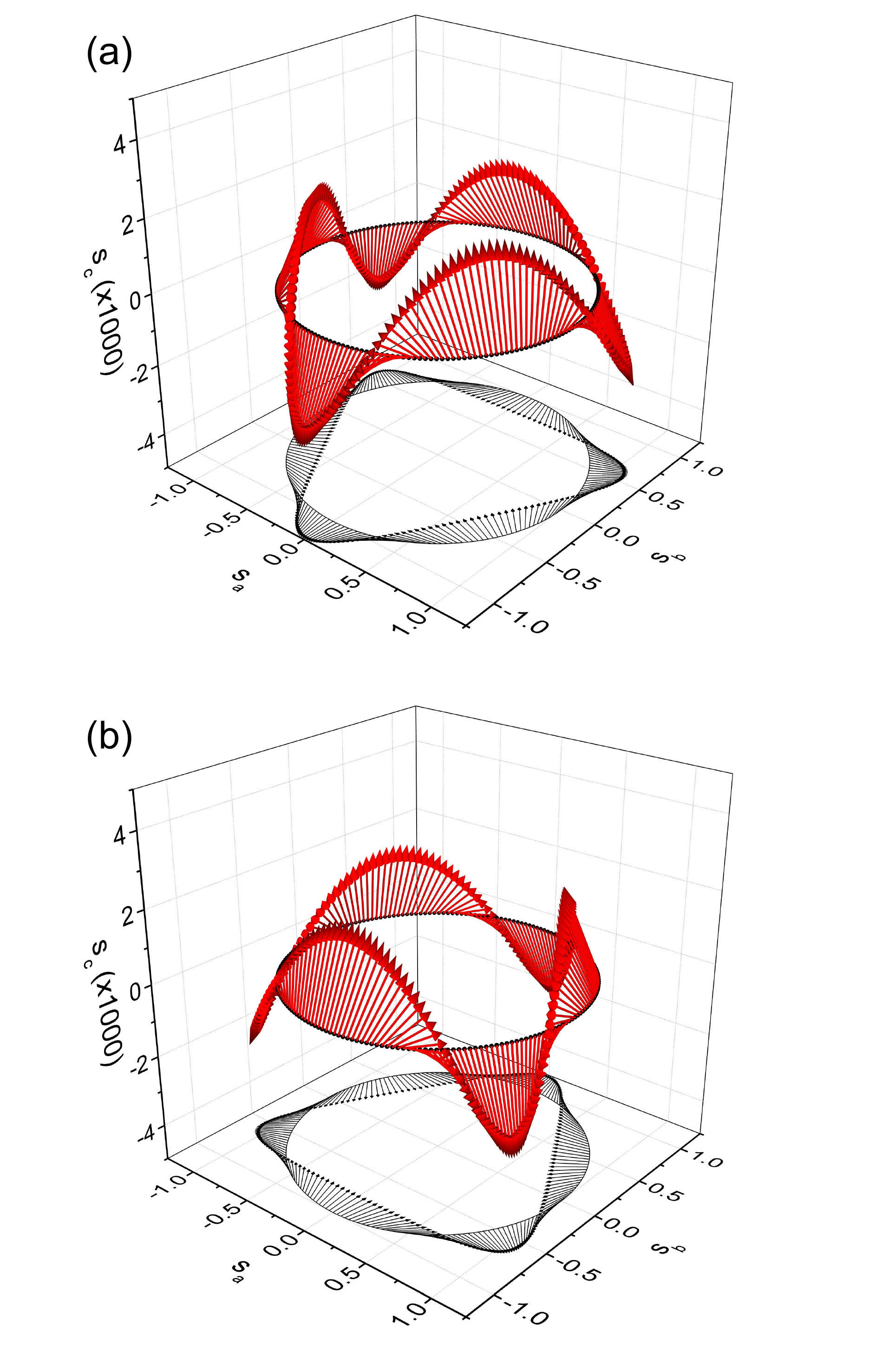}
	\caption[j=3/2]{Spin precession and nutation for $j=\pm\frac{3}{2}$, $l=\pm1$ and $kr_0=1$ for an InAs nanowire. In (a) and (b) the vector of the spin density of the upper and lower nondegenerate branches (cf.\ Fig.~\ref{fig:InAs-InSb-dispersion}) are depicted, respectively. The reference point of the spins follows $\varphi$ along the nanowire surface, as illustrated by the black circle. The vector components have been scaled with a factor of $0.25$. Black vectors represent the projection on the $ab$-plane.}
	\label{fig:sj3-2_l1}
\end{figure}

\begin{figure*}[htb]
	\centering
	\includegraphics[width=0.95\linewidth]{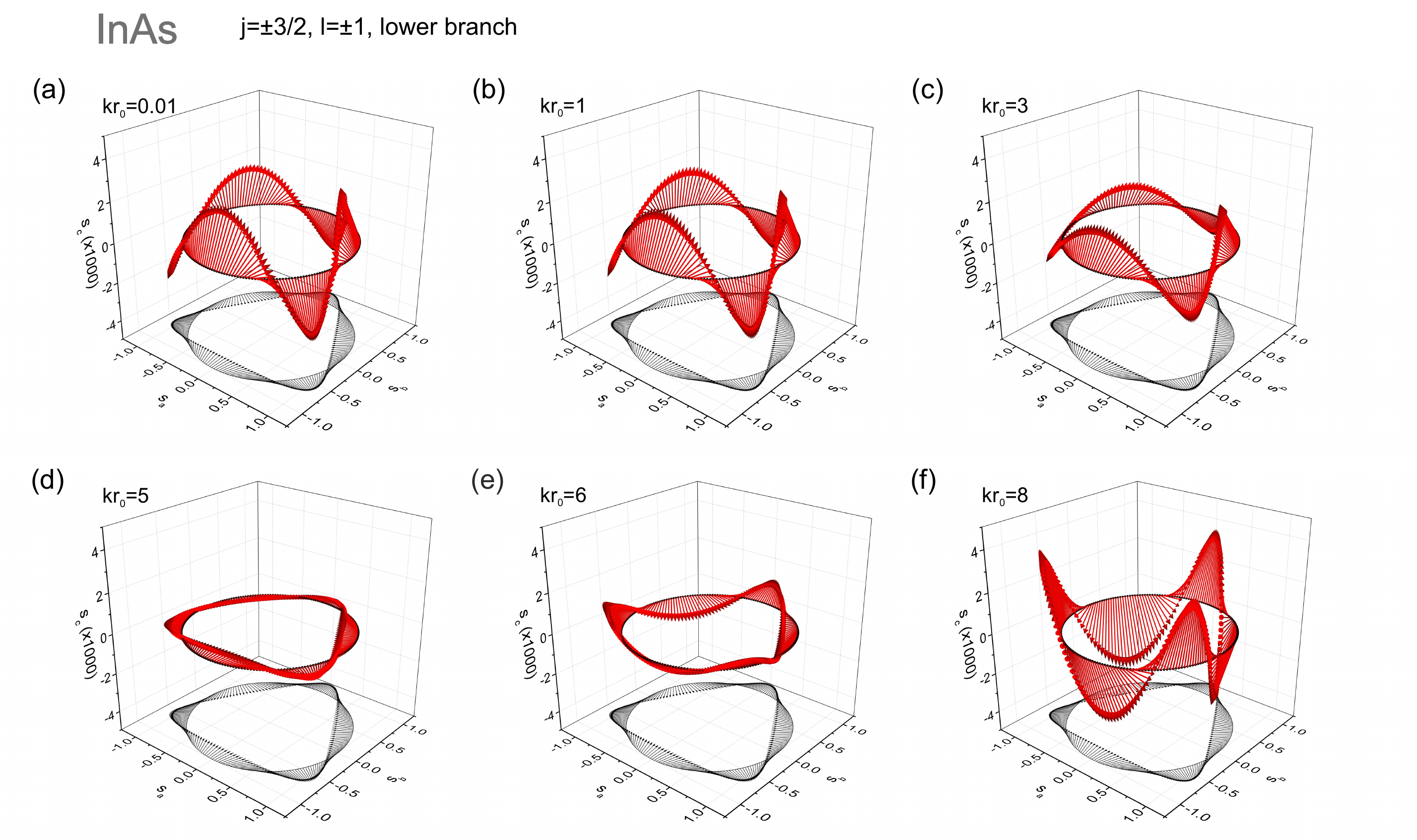}
	\caption[j=3/2]{Spin precession and nutation for $j=\pm\frac{3}{2}$, $l=\pm1$ for the lower branch for an InAs nanowire. From (a) to (f) the normalized momentum $kr_0$ was successively increased from $0.01$ to $8$. The components of the vectors have been scaled with a factor of $0.25$. Black vectors represent the projection on the $ab$-plane.}
	\label{fig:sj3-2_l1_k-dep_InAs}
\end{figure*}

We find that for the ``3/2" class the energetically lowest pair of nondegenerate modes is governed by a superposition of the $j=\pm\frac{3}{2}$, $l=\pm1$ states with equal weight. As a consequence, the spin expectation value along the nanowire axis is zero. However, this does not imply that the local spin density vanishes as well. The three components of the spin density $s_{a,b,c}$ follow from the spinor wave function $\Psi$: $s_{a,b,c} = \left(\Psi^\dagger \sigma_{a,b,c}\Psi\right)/2$. The standard representation of $\Psi$ is  $\Psi^{\uparrow} = f; \;  \Psi^{\downarrow} = g +i h$  with real functions of space $f$, $g$, and $h$. With these the components of the spin density are 
\begin{equation} s_a = fg , \enspace  s_b = fh , \enspace  s_c = \left( f^2 -g^2-h^2\right)/2 \, .
\end{equation}
In fact, as illustrated in Fig.~\ref{fig:sj3-2_l1}, we find that the spin density is strongly modulated. By integrating the spin density, one obtains that the spin expectation values in all three spatial directions vanish. Interestingly, by comparing the spin density for the upper and lower nondegenerate branches, as shown in Figs.~\ref{fig:sj3-2_l1}(a) and \ref{fig:sj3-2_l1}(b), respectively, one finds that the spin density pattern is rotated by an angle of $\pi/3$. Thus, an equal superposition of these states would result in a complete cancellation of the spin.
%%%%% Fig. 4
The spin texture changes with increasing linear momentum $\hbar k$, as can be seen in Fig.~\ref{fig:sj3-2_l1_k-dep_InAs} for the lower branch. Up to $kr_0=5$ the amplitude of the spin modulation decreases. Above $kr_0=5$ it increases again with reversed sign of the $c$-component of the spin.
%%%%% Fig. 5
 
\begin{figure*}[htb]
	\centering
\includegraphics[width=0.7\linewidth]{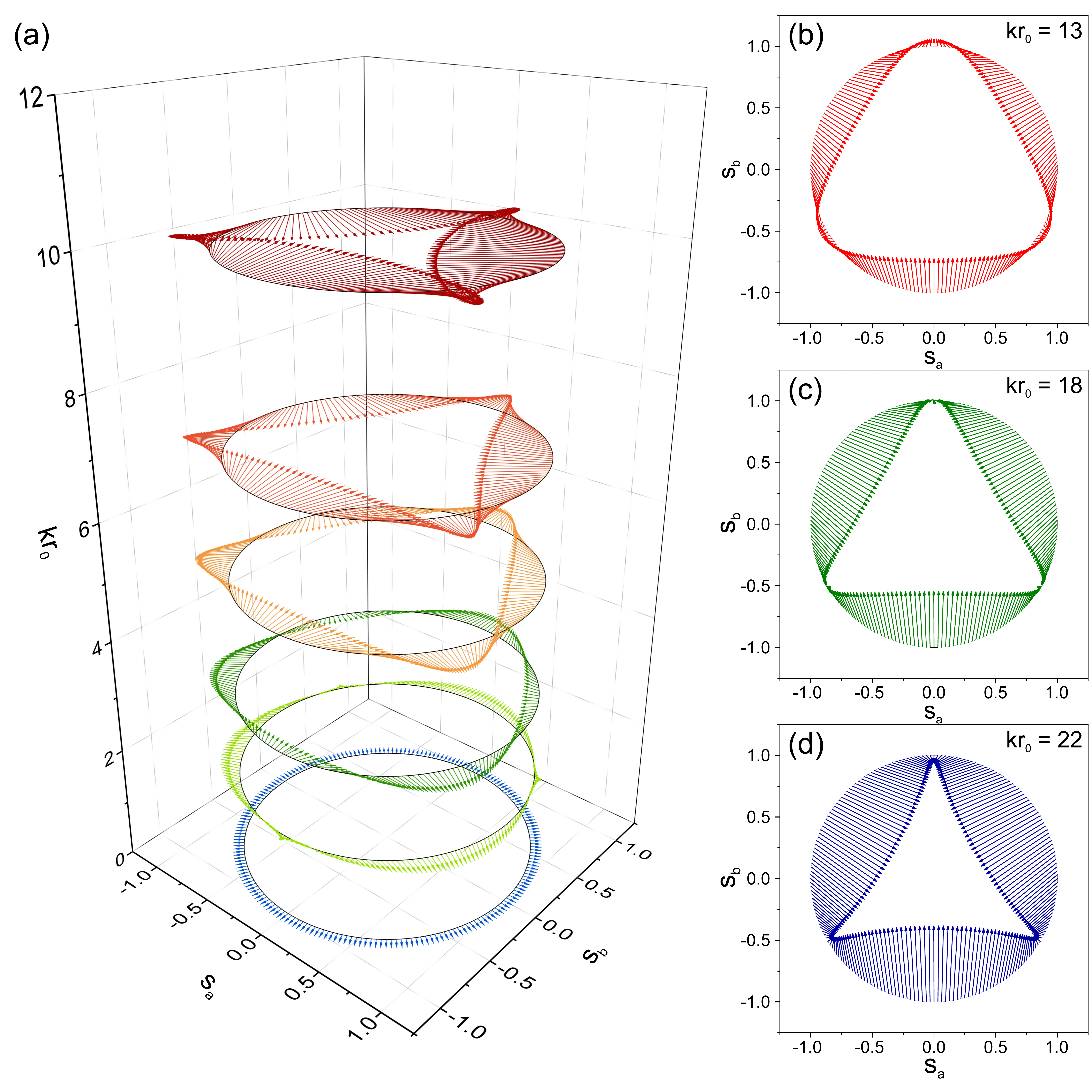}
     \caption[j=1/2]
         {(a) Spin texture of the $a$- and $b$-components for $j=\frac{1}{2}$, $l=0$ for $kr_0$ ranging from 0 to 10. The $c$-components have been omitted for clarity, since they dominate the spin direction. The vectors are scaled with a factor of $8$. The spin performs a nutation and precession, which become more pronounced with increasing axial momentum. (b) - (d) Corresponding projection of spin texture for $kr_0=13$, $18$, and $22$, respectively. Here, a lower scaling factor of $2$ was used.}
	\label{fig:sj1-2_l0}
\end{figure*}

While the states of the class``3/2" are characterized by the compensation of all spin-components, the states of the classes``1/2" and ``$-1/2$" have spin expectation values close to $\pm \frac{1}{2}$ along the $c$-direction. As can be seen in Fig.~\ref{fig:sj1-2_l0}, the spin density is modulated in the $ab$-plane for the lowest energy state ($j=\frac{1}{2}$, $l=0$). 
%%%%% Fig.6
The classically analog situation is a gyroscope under the influence of an external force. The characteristic motions a``precession'' -- rotation of the gyroscope-axis around the symmetry-axis -- and ``nutation'' -- periodic change of inclination of the gyroscope-axis against the symmetry-axis. The local spin direction  rotates, i.e.\, precedes, around the cylinder-axis and changes its $c$-component, i.e.\ nutates. This is illustrated in Fig.~\ref{fig:sj1-2_l0}(a) for increasing values of the normalized axial momentum $kr_0$. At $kr_0=0$ the spin component in the $ab$-plane points in the radial direction with a constant length, thus no nutation takes place. For non-zero values of $kr_0$, the spin component in the $ab$-plane is modulated in magnitude and deviates from a purely radial orientation, i.e.\ a nutation occurs in addition to the precession. The local spin periodically changes its orientation with respect to the precession axis. For $kr_0 \geq 4$ one even finds that the spin density winds around the precession axis. Compared to the $kr_0<4$ case, it changes the direction of rotation from a onefold turnaround counterclockwise to a twofold turnaround clockwise, i.e.\ the winding number is changed from $+1$ to $-2$. Around $kr_0=18$ the winding number once again changes from $-2$ to $+1$ when increasing the axial momentum. This can be seen in the series of spin projections shown in Figs.~\ref{fig:sj1-2_l0}(b) - (d). Compared to the case at zero momentum the spin orientation basically has an opposite phase.  

\section{Magnetic field effects}

The most eminent action of $H_\mathrm{D}$ is the compensation of angular momentum of the $j=\pm\frac{3}{2}$ states. It should be visible in their magnetic response. An axial magnetic field $B_c$ can be incorporated in the present calculation by including the diamagnetic and paramagnetic energy contributions to the Hamiltonian $H_0$ [cf.\ Eq.~\eqref{eq:H-null}] of the model system:
\begin{equation}
  H = H_0 + \frac{e^2}{2m^*}\,r^2B_c^2 + \mu_\mathrm{B}B_c \left(l+gs_c\right) \; . 
\label{eq:H-B}
\end{equation}
Here, $e$ is the electron charge, $\mu_\mathrm{B}$ the Bohr magneton, and $g$ the gyromagnetic factor (-14.92 in InAs and -51.56 in InSb~\cite{Winkler03}). The extra terms may be easily  incorporated into the perturbation procedure.

\begin{figure}[htb]
	\centering
\includegraphics[width=0.98\linewidth]{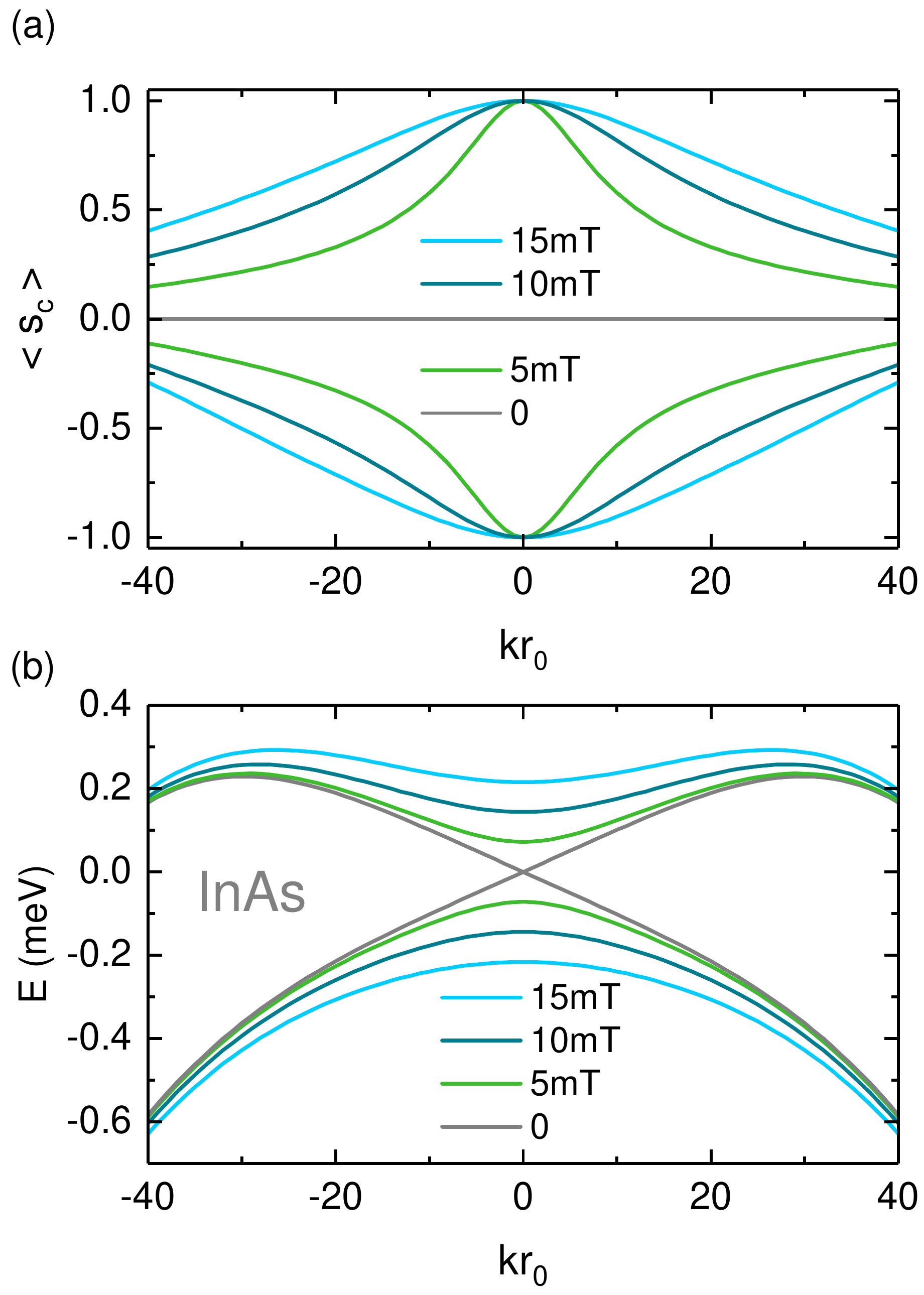}
	\caption[]{(a) Spin density $\langle s_c\rangle $ of InAs for $j=\pm3/2$, $l=\pm 1$ at various axial magnetic fields: $B=0,\,5,\,10,\,15$\,mT. (b) Corresponding energy vs. $k$ dispersion. }
	\label{fig:sj3-2_l1-Bf}
\end{figure}

At zero magnetic field the expectation values of the angular momenta $\langle j_c \rangle$ and $ \langle s_c \rangle $ are zero for "$j=\pm\frac{3}{2}$''-states except at $k=0$, where the two states are degenerate. A magnetic field removes this degeneracy via a Zeeman energy and results in a spin splitting, as shown in Fig.~\ref{fig:sj3-2_l1-Bf}(a). 
%%%%%%% Fig7
Since the coupling via the Dresselhaus interaction increases linearly with $k$, the spin expectation value $\langle s_c \rangle$ at a fixed $B_c$ eventually goes down to zero. Likewise, the magnetic contribution to the energy diminishes. Close to $k=0$ the states behave like Zeeman levels before the Dresselhaus interaction dominates, as shown in Fig.~\ref{fig:sj3-2_l1-Bf}(b).

Combining the dispersion given in Fig.~\ref{fig:sj3-2_l1-Bf}(b) with the kinetic energy contribution, one would in principle obtain the same shape of dispersion that results when combining the Rashba effect with an external magnetic field.\cite{Streda03} However, in the latter case, the Rashba effect leads to helical states. By covering the nanowire with a superconducting electrode the proximity effect results in the formation of Majorana zero modes.\cite{Lutchyn10,Oreg10} In our case the situation is different. At zero magnetic field we have no net spin polarization. Upon applying a magnetic field, the spins are aligned either parallel or antiparallel to the magnetic field. Thus, these states are nonhelical. Consequently, it is not expected that Majorana zero modes can be formed.  

The states of the classes ``1/2" and ``$-1/2$" are degenerate and show in general normal magnetic behavior. Nevertheless, as illustrated in Fig.~\ref{fig:InAs-dispersion-l-3-helical}(a), there are hybridization effects for states in these classes starting from a common $|l|=|j_c\pm s_c|$, e.g.\ $l=-3=-5/2-1/2$ and $l=3=7/2-1/2$ of class ``$-1/2$" or  $l=3=5/2+1/2$ and $l=-3=-7/2+1/2$ of class ``1/2". In both cases there is an interaction between the two states, as shown for InAs in  Fig.~\ref{fig:InAs-dispersion-l-3-helical}(a).  In order to discuss the effect of a magnetic field we restrict ourself to the state $j=+7/2$ ($l=+3,\,s=+1/2$) which is transfered to the state $j=-5/2$ ($l=-3,\, s=+1/2$) after passing the hybridization region. As illustrated in Fig.~\ref{fig:InAs-dispersion-l-3-helical}(b), the effect of reversing the orbital angular momentum from $+3$ to $-3$ becomes visible in the paramagnetic shift when a small magnetic field is applied. 

Upon applying a field of $10\,\mu$T a general downwards energy shift occurs owing to the dominating contribution of the spin originating from the paramagnetic term in Eq.~\eqref{eq:H-B} in connection with the large negative $g$ factor. However, for $l=+3$ at small values of $kr_0$ this downwards shift is reduced, while for $l=-3$ beyond $kr_0 \approx 3$ it is enhanced. The corresponding total energy shift $\Delta E= E(B_c=10\,\mu\mathrm{T})-E(B_c=0)$ is plotted in Fig.~\ref{fig:InAs-dispersion-l-3-helical}(c), solid line. The dashed line shows the energy shift for the initial state $j=-5/2$ ($l=-3, s=+1/2$). Once again, owing to the dominant contribution of the spin, the energy shift is downwards, however, the paramagnetic shift caused by the orbital angular momentum is opposite compared to the previous case. The effects discussed here disappear for magnetic fields much stronger than the Dresselhaus interaction. This interplay of spin-orbit interaction and magnetic field is well-known in atomic physics as the ``Paschen-Back'' effect, i.e.\ at low fields the level splitting is determined by the total angular momentum whereas at high fields by the spin alone.

\section{Conclusion}

In conclusion, the Dresselhaus contribution for InAs and InSb nanowires to the energy spectrum as well as  the corresponding spin distribution was calculated. We found that in contrast to bulk systems the geometrical confinement results in a weak but noticeable effect on the energy vs.\ momentum dispersion for $\vec{k} \parallel [111]$. However, for a subset of states belonging to a certain symmetry class even a relatively large energy splitting was found. For InSb a splitting in the order of meV is predicted for moderate values of the linear momentum $k$. Since these states originate from states with opposite spins, the net spin polarization is zero, although the spin density is modulated around the cylinder axis. For all other states the spin is mainly polarized along the nanowire axis.  Nevertheless, a small modulation of the spin density is present in the cross-sectional plane. Generally, the spin modulation depends on the linear momentum. 

Applying an axially oriented magnetic field has two effects. First, for the strongly coupled states of class ``3/2" with $l=\pm 1$, the contribution of the Zeemann effect lifts the degeneracy at $k=0$, i.e.\ resulting in an energy splitting. Furthermore, the initially unpolarized spin states get polarized. Interestingly, this polarization is weaker at larger $k$ values because the Dresselhaus contribution becomes dominant. Second, for the higher lying hybridizing levels the switching of the angular momentum results in a paramagnetic energy shift upon applying a magnetic field. 

Compared to the Rashba effect the Dresselhaus contribution is rather small. It might therefore be neglected when it comes to the formation of a helical gap in conjunction with Zeeman splitting \cite{Streda03,Heedt17}. This is in particular true for the lowest states with $j=\pm 1/2$ and $l=0$, since these states are the basis for creating Majorana fermions. However, if higher levels have to be taken into account the relatively strong splitting of the $j=\pm \frac{3}{2}$ states can be relevant. Nevertheless, we found that for an axial magnetic field no helical states are formed for $j=\pm \frac{3}{2}$, thus it is not expected that Majorana zero modes are formed. In this respect, it would be interesting to analyze the case with a magnetic field applied perpendicular to the nanowire axis.    

\section{Acknowledgment}

We thank Michael Kammermeier and Paul Wenk from University of Regensburg for fruitful discussions. This work was performed in collaboration with the Virtual Institute for Topological Insulators (VITI), which is financially supported by the Helmholtz Association.

\appendix

\section{Dresselhaus Hamiltonian in cylindrical coordinates}

The Dresselhaus Hamiltonian Eq.~\eqref{eq:Ham02} has three parts, $H_{\mathrm{D},1}$, $H_{\mathrm{D},2}$, and $H_{\mathrm{D},3}$ acting differently on the states of the electrons. With the representations of
spin- and momentum-operators given in Eq.~\eqref{eq:trafo} one finds for $H_{\mathrm{D},1}$ [first line of Eq.~\eqref{eq:Ham02}]:\\

     $$  k_a\sigma_b\,=\,\left(\cos\varphi\ \frac{1}{i}\partial_r
                                            -\sin\varphi\ \frac{1}{ri}\partial_\varphi\right)
                                    \frac{\sigma_+ -\sigma_-}{i} \; , $$
     $$  k_b\sigma_a\,=\,\left(\sin\varphi\ \frac{1}{i}\partial_r
                                           +\cos\varphi\ \frac{1}{ri}\partial_\varphi\right)
                                     \left(\sigma_+ +\sigma_-\right) \; ,  $$ 
     $$ k_a\sigma_b\,-\, k_b\sigma_a\,=\,-\sigma_+\,e^{-i\varphi}
                                    \left(\partial_r\,+\,\frac{\partial_\varphi}{ri}\right)
                                                                  +\sigma_-\,e^{i\varphi}
                                    \left(\partial_r\,-\,\frac{\partial_\varphi}{ri}\right) , $$ 
     $$k_a^2+k_b^2+k_c^2\,=\,-\nabla^2
                          \,=\, -\left(\partial_r^2\,+\,\frac{1}{r}\partial_r\,  
                                             +\,\frac{1}{r^2}\partial_\varphi^2\,+\,\partial_c^2\right) 
                                                  \; ,  $$
     \[ H_{\mathrm{D},1}\,=\,\frac{\gamma_{\mathrm{D}}}{2\sqrt{3}}
                                   \left[ -\sigma_+\,e^{-i\varphi}\,\mathfrak{D}_1
                                            +\sigma_-\,e^{i\varphi}\,\mathfrak{D}_2 \right] . \]
The differential operators  \\
$$  \mathfrak{D}_{1,2}\,=\,
       \left(\partial_r\,\pm\,\frac{1}{ri}\partial_\varphi\right)
         \left(\nabla^2-5\partial_{c}^2\right) $$
are of third order and act on the radial part of the wave function only. Linear- and angular-momenta and spin remain unchanged by $\mathfrak{D}_{1,2}$. $H_{\mathrm{D},1}$ turns the spin, but does not affect the total angular momentum $j$ because of the compensation effects between $\sigma_+$ and $e^{-i\varphi}$ and between $\sigma_-$ and $e^{i\varphi}$, respectively.\\
    
For $H_{\mathrm{D},2}$ (second line of Eq.~\eqref{eq:Ham02}) it follows from
  $$  2k_ak_bk_c\, =-\sin(2\varphi)\ \mathfrak{D}_3 -\cos(2\varphi)\ \mathfrak{D}_4
                                \; ,  $$
  $$  \left(k_a^2-k_b^2\right)k_c\,
              = -\cos(2\varphi)\ \mathfrak{D}_3 +\sin(2\varphi)\ \mathfrak{D}_4 \; , $$
  $$   \mathfrak{D}_3\,=\,\left(\partial_r^2\,  
                                     -\,\frac{1}{r}\partial_r\, 
                                     -\,\frac{1}{r^2}\partial_\varphi^2\right)
                                       \frac{1}{i}\partial_{c} \ ,\ 
          \mathfrak{D}_4\,=\, \partial_r\,\frac{2}{r}\,\partial_\varphi
                                       \frac{1}{i}\partial_{c}  \; ,  $$
 that the second part of the Hamiltonian can be written as
  $$ H_{\mathrm{D},2}\,=\,\frac{i\gamma_{\mathrm{D}}}{\sqrt{6}}
                                    \left[ e^{2i\varphi}\sigma_+ 
                                     \left(\mathfrak{D}_3+i\mathfrak{D}_4 \right)
                                  \,-\,e^{-2i\varphi}\sigma_-
                                        \left(\mathfrak{D}_3-i\mathfrak{D}_4 \right) \right].$$
It turns the spin and changes the total angular momentum $j$ to $j\pm 3$ because of the $e^{2i\varphi}\sigma_+$ and $e^{-2i\varphi}\sigma_-$ operators, respectively.

The operator in the third line of Eq.~\eqref{eq:Ham02} is:
   \[   \sigma_{c}\, k_{b}\,\left( k_{b}^2 -  3k_{a}^2\right)
          = \sigma_{c}\,i\,\,\Im \left(  k_{b} - i\,  k_{a} \right)^3   \]
The third power of $ k_{b} -i \,k_{a} $ is evaluated in polar coordinates
\begin{eqnarray} 
    &&\left( -\, e^{i\varphi}\partial_{r} -\frac{i}{r}  e^{i\varphi}\partial_{\varphi} \right)^3
             \nonumber \\
    &=&-\,e^{3i\varphi} \left( \partial_{r}^3\ -\ \partial_{r}\,\frac{1}{r}\,\partial_{r}
                            -\frac{2}{r}\partial_{r}^2+\frac{2}{r^2}\partial_{r}\right) \nonumber \\
    && -\,i\,\ e^{3i\varphi}  \partial_{\varphi}
           \left(\partial_{r}^2\,\frac{1}{r}+\partial_{r}\frac{1}{r}\,\partial_{r}
                 +\frac{1}{r}\partial_{r}^2 \right) \nonumber \\
    &&+\,i\, e^{3i\varphi}  \partial_{\varphi}
                \left(\partial_{r}\frac{1}{r^2}+\frac{2}{r}\partial_{r}\frac{1}{r}
                       +\frac{3}{r^2}\partial_{r}-\frac{2}{r^3} \right)\  \nonumber \\
    &&+\ e^{3i\varphi}  \partial_{\varphi}^2
                \left(\partial_{r}\frac{1}{r^2}\,+\,\frac{1}{r}\,\partial_{r}\,\frac{1}{r}
                       +\frac{1}{r^2}\partial_{r}-\frac{3}{r^3}\right) \nonumber  \\
    &&+ \ i\,\ e^{3i\varphi} \left( \partial_{\varphi}/r \right)^3 \; .  \nonumber
\end{eqnarray}
After carrying out the differentiations with respect to $r$ one gets:
\[   H_{\mathrm{D},3}\,=\, -\frac{i\gamma_{\mathrm{D}}}{\sqrt{6}}\,\sigma_{c}
                                   \left[ \sin(3\varphi)\,\mathfrak{D}_5\,+\,
                                             \cos(3\varphi)\,\mathfrak{D}_6 \right] \]
with the differential operators given by  
    $$   \mathfrak{D_5}\,=\,
                                         \partial_{r}^3 - \frac{3}{r}\partial_{r}^2
                                         + \frac{3}{r^2}\partial_{r}
                                       -  \left( \frac{3}{r^2}\partial_{r}-\frac{6}{r^3}\right)
                                                 \,\partial_{\varphi}^2. $$
    $$   \mathfrak{D_6}\,=\,
                                        \left( \frac{3}{r}\partial_{r}^2 -\frac{9}{r^2}\partial_{r}
                                                + \frac{8}{r^3}\right)\,\partial_{\varphi} 
                                                - \left( \frac{\partial_{\varphi}}{r} \right)^3  $$
 $H_{\mathrm{D},3}$ changes $j$ to $j \pm 3$ and conserves the spin.

\section{Recursion relation}

In total, $H_{\mathrm{D}}$ connects only states $\chi_{n,j}^{\sigma}$ (cf.\ Eqs.~\eqref{eq:chi-up} and \eqref{eq:chi-down}), with $j$ differing by $\pm 3$ ($\sigma=\uparrow,\downarrow$). An expansion of stationary states $\Psi$ with respect to $\chi_{n,j}^{\sigma}$:
  $$
\centerline{  $\Psi\,=\,\sum_{n,j,\sigma} \Gamma_{n,j}^{\sigma}\,\chi_{n,j}^{\sigma}$} 
  $$
leads to a recursive system of equations for the coefficients $\Gamma$ :
\begin{eqnarray}
   0 &=&   \left(\varepsilon_{k,n,l}-{\cal E}\right)
                            \Gamma_{n,j}^{\uparrow} \nonumber \\
      &+&  \sum_{n'} \left[ \,\langle\chi_{n,j}^{\uparrow}\,|\,H_{\mathrm{D},1}\,|\chi_{n',j}^{\downarrow} \rangle \, \Gamma_{n',j}^{\downarrow}  \right. \nonumber  \\
      &+&  \qquad \left.                \langle\chi_{n,j}^{\uparrow}\,|\,H_{\mathrm{D},2}\,|\chi_{n,'j-3}^{\downarrow} \rangle \, \Gamma_{n',j-3}^{\downarrow} \right.  \nonumber \\
      &+&  \qquad \left. \langle\chi_{n,j}^{\uparrow}\,|\,H_{\mathrm{D},3}\,|\chi_{n',j-3}^{\uparrow} \rangle \, \Gamma_{n',j-3}^{\uparrow} \right. \nonumber \\
      &+& \qquad \left.  \langle\chi_{n,j}^{\uparrow}\,|\,H_{\mathrm{D},3}\,|\chi_{n',j+3}^{\uparrow} \rangle \, \Gamma_{n',j+3}^{\uparrow} \right]   \nonumber
\end{eqnarray}
with $l=j-1/2$, and with $l=j+1/2$
\begin{eqnarray}
    0 &=&   \left(\varepsilon_{k,n,l}-{\mathcal{E}}\right)
                           \Gamma_{n,j}^{\downarrow} \nonumber \\
       &+&  \sum_{n'} \left[
       \,\langle\chi_{n,j}^{\downarrow}\,|\,H_{\mathrm{D},1}\,|\chi_{n',j}^{\uparrow} \rangle \, \Gamma_{n',j}^{\uparrow}  \right. \nonumber  \\
       &+& \qquad \left. \langle\chi_{n,j}^{\downarrow}\,|\,H_{\mathrm{D},2} \,|\chi_{n,'j+3}^ \uparrow \rangle  \, \Gamma_{n',j-3}^{\uparrow} \right.  \nonumber \\
       &+& \qquad \left.\langle\chi_{n,j}^{\downarrow}\,|\,H_{\mathrm{D},3}\,|\chi_{n',j-3}^{\downarrow} \rangle \, \Gamma_{n',j-3}^{\downarrow} \right. \nonumber \\
       &+& \qquad \left.\langle\chi_{n,j}^{\downarrow}\,|\,H_{\mathrm{D},3}\,|\chi_{n',j+3}^{\downarrow} \rangle \, \Gamma_{n',j+3}^{\downarrow} \right]  \nonumber \; . 
\end{eqnarray}
This system can be solved by back-folding from large to small values of $|j|$ and leads to a convergent series for $\Psi$. It is sufficient to take the basis states with $n$ and $\,n'$ equal to 0 and 1 into account.

%\newpage
%\bibliography{Dresselhaus-SOI-references}
%\newpage

%\end{document}

\newpage

\end{document}